# FairFML: Fair Federated Machine Learning with a Case Study on Reducing Gender Disparities in Cardiac Arrest Outcome Prediction

Siqi Li, BSc[1#], Qiming Wu, MSc[1#], Xin Li, MSc[1], Di Miao, MSc[1], Chuan Hong, PhD[2], Wenjun Gu, BSc[1], Yuqing Shang, MSc[1], Yohei Okada, PhD[3,4], Michael Hao Chen, BSc[1], Mengying Yan, MSc[2], Yilin Ning, PhD[1], Marcus Eng Hock Ong, MPH[3,5,6], Nan Liu, PhD[1,3,7]*

[1] Centre for Quantitative Medicine, Duke-NUS Medical School, Singapore, Singapore

[2] Department of Biostatistics and Bioinformatics, Duke University, Durham, NC, USA

[3] Programme in Health Services and Systems Research, Duke-NUS Medical School, Singapore, Singapore

[4] Department of Preventive Services, Graduate School of Medicine, Kyoto University, Kyoto, Japan

[5] Health Services Research Centre, Singapore Health Services, Singapore, Singapore

[6] Department of Emergency Medicine, Singapore General Hospital, Singapore, Singapore

[7] NUS Artificial Intelligence Institute, National University of Singapore, Singapore, Singapore

[#] These authors contributed equally to this work.

* Correspondence: Nan Liu, Centre for Quantitative Medicine, Duke-NUS Medical School, 8 College Road, Singapore 169857. Phone: +65 6601 6503. Email: liu.nan@duke-nus.edu.sg

# Abstract

**Objective**

Mitigating algorithmic disparities is a critical challenge in healthcare research, where ensuring equity and fairness is paramount. While large-scale healthcare data exist across multiple institutions, cross-institutional collaborations often face privacy constraints, highlighting the need for privacy-preserving solutions that also promote fairness.

**Materials and Methods**

In this study, we present Fair Federated Machine Learning (FairFML), a model-agnostic solution designed to reduce algorithmic bias in cross-institutional healthcare collaborations while preserving patient privacy. As a proof of concept, we validated FairFML using a real-world clinical case study focused on reducing gender disparities in cardiac arrest outcome prediction.

**Results**

We demonstrate that the proposed FairFML framework enhances fairness in federated learning (FL) models without compromising predictive performance. Our findings show that FairFML improves model fairness by up to 65% compared to the centralized model, while maintaining performance comparable to both local and centralized models, as measured by receiver operating characteristic analysis.

**Discussion and Conclusion**

FairFML offers a promising and flexible solution for FL collaborations, with its adaptability allowing seamless integration with various FL frameworks and models, from traditional statistical methods to deep learning techniques. This makes FairFML a robust approach for developing fairer FL models across diverse clinical and biomedical applications.

# 1. Introduction

Machine learning (ML) and artificial intelligence (AI) methods have been rapidly adopted in healthcare for predictive modeling, offering substantial potential to improve patient outcomes[1]. However, ensuring health equity–an essential principle in population and public health research[2]–remains a critical challenge, particularly when algorithmic findings directly impact clinical decision-making and patient care[3]. Growing concerns have been raised about the potential for ML and AI systems to underperform in historically underserved populations, including women and individuals from lower socioeconomic backgrounds[4]. In response, researchers have developed various qualitative and quantitative approaches to promote fairness in clinical model development[5].

Algorithmic disparity[6], commonly referred to as "biased" or "unfair" decision-making, arises when predictive models perform unequally across subgroups[7,8] defined by sensitive attributes such as gender, race/ethnicity, and socioeconomic status[5]. For instance, studies have shown that Black patients are more frequently underdiagnosed with chronic obstructive pulmonary disease (COPD) compared to Hispanic white patients[4,9], highlighting the importance of addressing these disparities when defining COPD prevalence and improving population health[9]. This is just one of many documented inequities spanning healthcare domains such as COVID-19[3,6,7], stroke[10], emergency medicine[11–13], cardiovascular disease[14], cancer[15], and organ transplants[16].

Despite growing efforts to develop fair models[5], most studies rely on single, centralized datasets. However, healthcare data are often distributed across multiple sites, such as electronic health records (EHRs) from different hospitals or mobile health data from users' devices[17,18]. Aggregating these diverse data sources could accelerate research and improve care quality[19], but privacy regulations pose significant barriers[20]. Federated learning (FL) or federated ML (FML), an ML technique, enables participants (i.e., clients) to collaboratively train models without exchanging data[18], making it an increasingly popular approach in medical research[21,22].

While FL adoption is increasing, most studies focus primarily on overall predictive performance, often overlooking its potential to address algorithmic disparities[22]. Evidence suggests that standard FL algorithms struggle to reduce algorithmic biases[23,24], leading to models that retain their unfairness when transitioning from single-site analyses to FL settings. Although some studies have investigated these disparities within FL contexts, they predominantly rely on conventional ML datasets rather than real-world clinical data[24–26], raising concerns about the generalizability of their findings to actual healthcare settings.

To address these gaps, we propose Fair Federated Machine Learning (FairFML), a unified solution to promote fairness in FL. As a proof of concept, we use real-world out-of-hospital cardiac arrest (OHCA) data from the United States, focusing on gender disparities—a critical concern for equity in OHCA care—to demonstrate FairFML's effectiveness[12,27,28]. This case study shows that FairFML enables participating sites to develop fairer models for sensitive groups while maintaining prediction performance comparable to both local and centralized analyses.

## 2. Materials and Methods

### 2.1 Notation and problem setup

In this study, we adopt the notation introduced by Berk et al.[29]. Let $y \in \mathcal{Y} = [-1,1]$ represent the binary outcome, and $x \in \mathcal{X} = R^d$ denote the feature vectors. Each instance is categorized into one of two groups based on a sensitive variable, denoted as $\mathcal{X}_{d+1}$. The joint distribution of $\mathcal{X}$ and $\mathcal{Y}$ is represented by $\mathcal{P}$. We consider a training set $S = \{(x_i, y_i)\}_{i=1}^{n}$, consisting of $n$ independent and identically distributed (i.i.d.) samples drawn from $\mathcal{P}$. This training set is divided into two groups, $S_1$ and $S_2$, based on the sensitive variable, with $n_1$ and $n_2$ representing the respective sizes of these groups, such that $n_1 + n_2 = n$.

The λ-weighted fairness loss for a given model is defined as $\mathcal{L}(w, S) + \lambda f(w, S)$, where $\mathcal{L}$ represents the standard model loss function, $w$ represents model parameters, and $\lambda$ is a regularization parameter for the fairness penalty. Consistent with Berk et al.https://www.zotero.org/google-docs/?t7WY1i[29], we focus on a group fairness penalty, defined as

$$f(\mathbf{w}, S) = \frac{1}{n_1 n_2} \sum_{\substack{(x_i, y_i) \in S_1 \\ (x_j, y_j) \in S_2}} d(y_i, y_j)(\mathbf{w} \cdot \mathbf{x}_i - \mathbf{w} \cdot \mathbf{x}_j)$$

Here, $d(y_i, y_j) = \mathbb{1}[y_i = y_j]$ serves as the cross-group fairness weight.

### 2.2 Group fairness metrics

Demographic parity (DP), also known as statistical parity, and equalized odds (EO) are two widely used algorithmic fairness definitions for binary classifications:

- A model satisfies DP over a distribution $\mathcal{P}$ if its prediction $\hat{Y}$ is statistically independent of the sensitive feature:

$$P[\hat{Y} = 1 \mid \mathcal{X}_{d+1} = a] = P[\hat{Y} = 1], \forall a$$

- A model satisfies EO over a distribution $\mathcal{P}$ if its prediction $\hat{Y}$ is conditionally independent of the sensitive feature given the true outcome label:

$$P[\hat{Y} = 1 \mid \mathcal{X}_{d+1} = a, Y = y] = P[\hat{Y} = 1 \mid Y = y], \forall a, y$$

In this study, we focused on a total of four fairness metrics: demographic parity difference (DPD), demographic parity ratio (DPR), equalized odds difference (EOD), and equalized odds ratio (EOR), which are calculated using the definitions of DP and EO as follows:

- DPD $= \max_{a} E[\hat{Y} \mid \mathcal{X}_{d+1} = a] - \min_{a} E[\hat{Y} \mid \mathcal{X}_{d+1} = a]$

- DPR $= \dfrac{\min_{a} E[\hat{Y} \mid \mathcal{X}_{d+1} = a]}{\max_{a} E[\hat{Y} \mid \mathcal{X}_{d+1} = a]}$

- EOD $= \max_{y \in \{-1,1\}} \left( \max_{a} E[\hat{Y} \mid \mathcal{X}_{d+1} = a, Y = y] - \min_{a} E[\hat{Y} \mid \mathcal{X}_{d+1} = a, Y = y] \right)$

- EOR $= \min_{y \in \{-1,1\}} \dfrac{\min_{a} E[\hat{Y} \mid \mathcal{X}_{d+1} = a, Y = y]}{\max_{a} E[\hat{Y} \mid \mathcal{X}_{d+1} = a, Y = y]}$

Lower values for DPD and EOD, and higher values for DPR and EOR, indicate greater fairness.

## 2.3 FairFML

We integrate the λ-weighted fairness loss described in Section 2.1 into the FL model training, and the workflow of our proposed FairFML is illustrated in Figure 1. As shown, incorporating FairFML into any FL framework enhances the fairness of existing FL solutions by replacing the standard model loss function $\mathcal{L}$ with the $\lambda$-weighted fairness loss function during FL model training. Since the fairness regularizer $f$ is convex[29], the combined objective function $\mathcal{L}(w, S) + \lambda f(w, S)$ remains convex as long as the standard loss function of the model is also convex. This convexity enables efficient optimization within most existing FL frameworks, such as FedAvg[30]. To prevent overfitting, we incorporate $l_2$ regularization, resulting in the final loss function: $\mathcal{L}(w, S) + \lambda f(w, S) + \gamma \|w\|_2^2$.

The trade-off between model accuracy and fairness, regulated by $\lambda$, varies significantly across datasets[29,31], where higher $\lambda$ values impose greater fairness penalties. As $\lambda$ increases from 0 to $\infty$, model accuracy tends to decrease. Therefore, users need to select an appropriate $\lambda$ value for each dataset to balance improved fairness with an acceptable reduction in model accuracy. To address this challenge, we propose a data-driven approach for efficiently selecting $\lambda$ while minimizing computational costs. As outlined in the pseudocode (eFigure 1, Supplementary Materials), $\lambda_k$ is initially chose independently for each client $k$ by plotting prediction metrics (e.g., accuracy or mean square error (MSE)) against $\lambda_k$. A practical method involves incrementing $\lambda_k$ in fixed steps until the prediction metrics degrade beyond a set threshold compared to the unregularized model (e.g., when accuracy falls below $0.995*\text{Acc}_0$, where $\text{Acc}_0$ is the accuracy of the model without the fairness penalty). The maximum $\lambda_k$ across all clients is then used to define the range for FL training, from which a user-defined set of equally spaced $\lambda$ values is selected.

For each $\lambda$ value, we use a two-step strategy to determine the optimal $\gamma$. First, we explore a broad, equally spaced $\gamma$ values starting from zero. The user selects the best $\gamma$ based on changes in predictive performance and fairness metrics. We then narrow the search range around that value and repeat the process to finalize $\gamma$ for the given λ. Detailed pseudocode for selecting $\gamma$ is provided in eFigure 1.

## 2.4 Dataset and Experiments

Our study population comprised OHCA patients treated by emergency medical services (EMS) providers, as recorded in the Resuscitation Outcomes Consortium (ROC) Cardiac Epidemiologic Registry (Epistry) (Version 3, covering the period from April 1, 2011, to June 30, 2015). The ROC, a North American database established in 2004, aims to advance clinical research on cardiopulmonary arrest[32]. Ethical approval was obtained from the National University of Singapore Institutional Review Board (IRB), which granted an exemption for this study (IRB Reference Number: NUS-IRB-2023-451).

We included patients aged 18 and older who were transported by EMS, achieved return of spontaneous circulation (ROSC) at any point prehospital, and had complete data on gender, race, etiology, initial rhythm, witness status, response time, adrenaline use, and neurological status. The primary outcome was neurological status at discharge, measured by the Modified Rankin Scale (MRS), where scores of 0, 1, or 2 were classified as a good outcome. Variables used for outcome prediction included age (in years), etiology of arrest (cardiac/non-cardiac), witness presence (yes/no), initial rhythm (shockable/non-shockable), bystander cardiopulmonary resuscitation (CPR) (yes/no), response time (in minutes), and adrenaline use (yes/no).

We conducted four sets of experiments to simulate real-world cross-site data by partitioning the full study cohort heterogeneously: (I) by race/ethnicity into four sites, (II) by age into four sites,

(III) by race/ethnicity into six sites, and (IV) by age into six sites. Continuous variables were standardized using the mean and standard deviation from the full cohort, and logistic regression was employed for outcome prediction. We focused on two representative FL frameworks, FedAvg and Per-FedAvg[33]. FedAvg is a foundational framework and the first proposed in the FL domain[30,34], while Per-FedAvg is a widely adopted solution for personalized FL. The latter is particularly relevant in healthcare data analysis, as it allows researchers to determine whether FL can offer localized benefits that enhance the performance of existing models for individual institutions[22].

For each scenario, we conducted three types of analyses: (1) a central model trained on the full cohort and local models trained independently at each site, (2) federated logistic regression using FedAvg and Per-FedAvg, and (3) fairness-enhanced federated logistic regression using the proposed FairFML method with the two FL frameworks–FairFML (FedAvg) and FairFML (Per-FedAvg). We evaluated model performance using the area under the receiver operating characteristic curve (AUROC) and four fairness metrics, as described in Section 2.2, with gender as the sensitive variable, using the 'Fairlearn' package[35].

## 3. Results

Figure 2 illustrates the partitioning of 7,425 individual episodes into four or six sites following the cohort formation process, with a 7:3 split for training and testing data. eTable 1 in the Supplementary Materials summarizes the baseline characteristics of the overall cohort and each site under different experimental conditions. In cases I and III, where clients were partitioned by race/ethnicity, significant distribution differences were observed, with the proportion of White individuals ranging from 88.9% to 48.8%. In cases II and IV, where clients were partitioned by age, the mean age varied considerably, ranging from approximately 80 to 60 years. Outcome prevalence varied from 7.5% to 12.6%, and other variables also exhibited heterogeneous distributions, reflecting the real-world demographic differences across regions.

Details of the experimental setup, including the tuning of $\lambda$ and $\gamma$ and other general hyperparameters for FL, are provided in eFigure 2 and eTable 2 in the Supplementary Materials. We assessed the performance of the federated model developed using FairFML by comparing it to the centralized model, local models trained independently at each site, and general FL models (FedAvg and Per-FedAvg). Figure 3 presents the average performance of each model across the testing datasets for all sites in each experimental scenario, with the percentage change in fairness metrics relative to the centralized model, displayed above each bar. Detailed results for each model and site are available in eTable 3, showing that at the client level, the fairness metrics generally improved across all clients, consistent with the average results across sites, with only a minor trade-off in prediction performance.

Key findings from Figure 3 and eTable 3 include: 1) FairFML consistently outperformed other models in fairness, improving all four fairness metrics by up to 65% compared to the centralized model, while maintaining predictive performance nearly identical to other models, with a maximum decrease in AUC of less than 0.02 relative to the centralized model; and 2) although FedAvg and Per-FedAvg occasionally outperformed central and local models on specific fairness metrics for certain clients, their improvements were less substantial. In contrast, FairFML consistently demonstrated significant and superior performance across all fairness metrics.

## 4. Discussion

FairFML offers a unified, model- and framework-agnostic solution[22,34] for enhancing fairness in FL collaborations. Its adaptability to various FL frameworks and ML models—ranging from traditional statistical regressions and support vector machines to deep neural networks—makes it highly versatile for clinical and biomedical prediction tasks[22]. By reducing algorithmic disparities, as demonstrated in our proof-of-concept case study focusing on gender disparities in cardiac arrest outcome prediction, FairFML effectively mitigates algorithmic bias when integrated with standard FL frameworks.

Given that clients in cross-institutional FL collaborations often expect direct benefits for their research or clinical practice[19,22], it is essential to evaluate models against both client-level (local) and central models. Our results show that FairFML consistently outperforms traditional FL and local models in terms of fairness between the two genders, as seen in Figure 3 and eTable 3. While previous studies have highlighted Per-FedAvg's ability to enhance client-level predictions through meta-learning-aided personalization[33], our findings suggest that FairFML (FedAvg) sometimes achieves better fairness improvements than FairFML (Per-FedAvg). This suggests that fairness improvements by FairFML are not exclusively dependent on the FL framework used. Thus, FairFML has strong potential for delivering enhanced fairness, with opportunities for further refinement in balancing fairness and client-level performance.

Although gender disparities in cardiac arrest are a key focus, they are not the only relevant partition for group fairness in this context[36]. Studies show that individuals from Black, Hispanic, or lower socioeconomic status backgrounds experience pronounced disparities throughout the resuscitation pathway[37]. Our findings, presented in eTable 4 of the Supplementary Materials, highlight significant variations in gender disparities across racial and ethnic groups.This underscores the need to address multi-group fairness (i.e., multiple intersecting sensitive variables[38]) to further mitigate unfairness. Despite more than a decade of discussion on multi-group fairness[39,40], it has received limited attention in FL settings. The challenge becomes more complex when group partitions are imbalanced or, in extreme cases, when certain groups are entirely absent from some clients. In such cases, advanced approaches, such as synthetic data[41,42], may offer the promising direction for future research.

Fairness in FL is typically framed around client resource allocation and ensuring performance uniformity across clients[43,44], commonly referred to as "system fairness"[45]. This is particularly relevant in scenarios involving client selection to optimize convergence speed and reduce computational costs[46], as seen in cross-device FL[18]. However, cross-institutional FL[18]—which is more prevalent in healthcare settings and often involves fewer clients (typically fewer than five)[22] —the focus shifts to algorithmic fairness. While various strategies have been proposed to enhance fairness in clinical models, including privacy-preserving collaborations, McCradden et al.[47] caution that relying solely on technical solutions may inadvertently harm vulnerable groups. Thus, FairFML should be viewed as a starting point, followed by further analysis of downstream patient impacts, rather than assuming that fairness can be achieved solely through ML/AI metrics[47].

**Limitations**

Our clinical case study uses simulated partitioned clients for FL experiments as a proof of concept, given the lack of suitable collaborators for a real-world setup. Although we simulate cross-site data heterogeneity, real-world collaborations may introduce additional complexities, particularly regarding model heterogeneity[22,34]. Further research is required to validate FairFML's robustness and applicability in real-world cross-institutional collaborations.

**Future work**

While this study focused on group fairness, our proposed method can be extended to improve individual fairness[48] by incorporating an individual fairness penalty within the convex framework[29]. A hybrid penalty combining both group and individual fairness metrics could offer a more comprehensive approach to mitigating unfairness in clinical research. Future work aims to explore these extensions and validate FairFML in real-world settings to ensure its robustness and applicability across diverse clinical environments.

## 5. Conclusion

FairFML effectively mitigates bias and enhances fairness in model co-training across multiple healthcare data owners while preserving privacy. In our proof-of-concept case study using real-world emergency medicine data, FairFML consistently outperformed other models in addressing fairness disparities without compromising predictive performance. These findings highlight FairFML's robustness and practical applicability to heterogeneous clinical data, making it a promising solution for real-world healthcare settings.

# Author Contribution

**Siqi Li:** Conceptualization, Project administration, Supervision, Method design, Algorithm development, Formal analysis, Data curation, Writing – original draft. **Qiming Wu:** Algorithm development, Formal analysis, Software, Writing – original draft. **Xin Li:** Data analysis, Writing – original draft. **Di Miao:** Formal analysis, Writing – original draft. **Chuan Hong:** Algorithm development, Writing – review & editing. **Wenjun Gu:** Data curation, Investigation, Writing – review & editing. **Yohei Okada:** Investigation, Validation, Writing – review & editing. **Michael Hao Chen:** Investigation, Validation, Writing – review & editing. **Mengying Yan:** Investigation, Validation, Writing – review & editing. **Yuqing Shang:** Algorithm development, Investigation, Writing – review & editing. **Yilin Ning:** Investigation, Validation, Writing – review & editing. **Marcus Eng Hock Ong**: Investigation, Validation, Writing – review & editing. **Nan Liu:** Conceptualization, Supervision, Funding acquisition, Resources, Writing – review & editing.

# Funding

This work was supported by the Duke/Duke-NUS Collaboration grant. The funder of the study had no role in the study design, data collection, data analysis, data interpretation, or writing of the report.

# Competing Interest

NL, SL and MEHO hold a patent related to the federated scoring system. The other authors declare no competing interests.

# Code Availability

The Python code for FairFML is available at https://github.com/nliulab/FairFML.

**Figure 1**. Workflow of FairFML.

**Figure 2**. Cohort formation flow diagram.

**Figure 3**. Performance comparison of the proposed FairFML method with baseline models, using gender as the sensitive variable.

**Figure 1**. Workflow of FairFML. Clients collaboratively train models using any FL framework, which may lead to algorithmic biases. FairFML mitigates these biases by enhancing fairness in the resulting models while remaining fully compatible with existing FL frameworks.

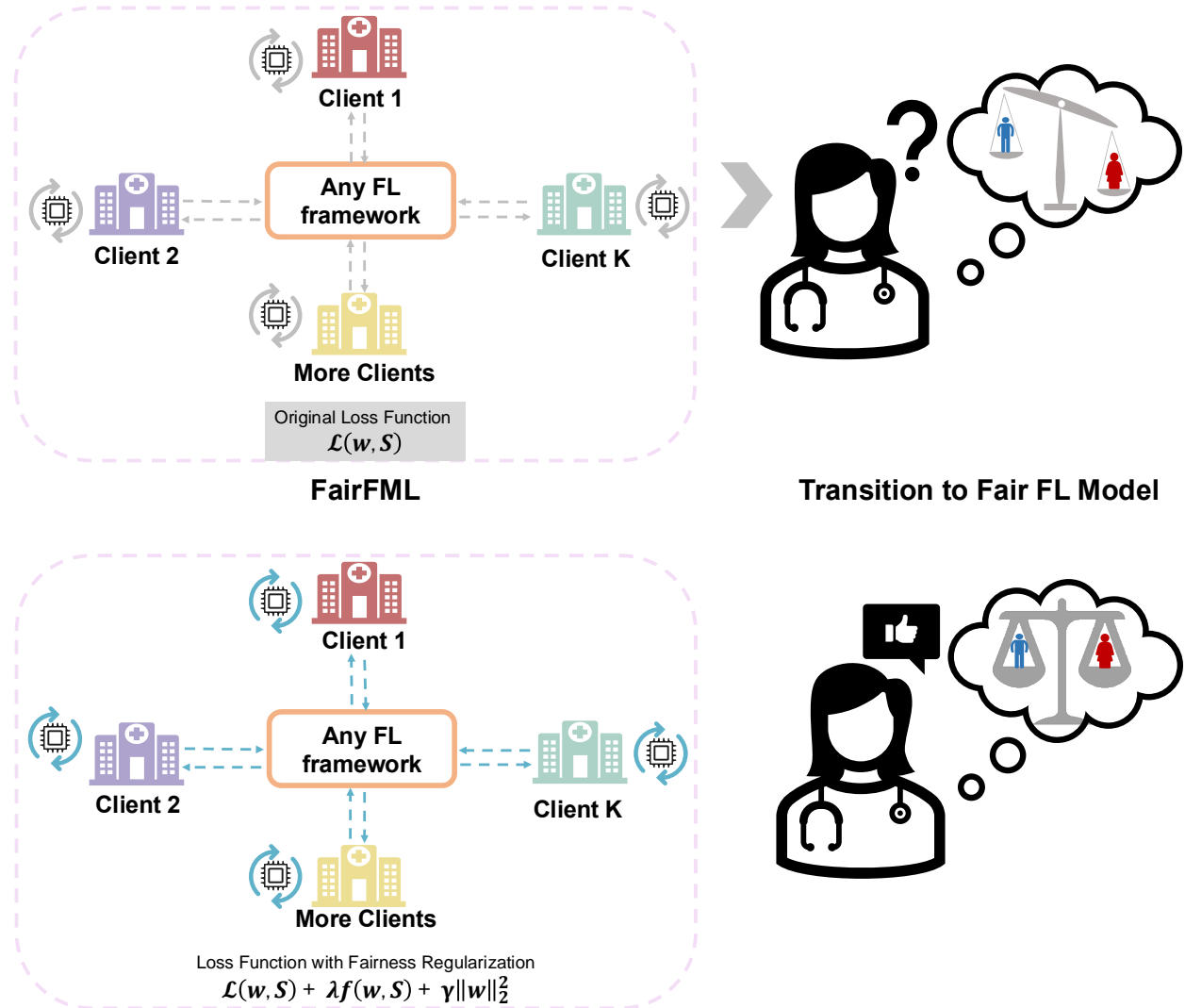

**Figure 2.** Cohort formation flow diagram. A total of 7,425 episodes were partitioned heterogeneously across clients by race/ethnicity (Cases I and III) and by age (Cases II and IV), with 70% of the data used for training and 30% for testing. Cases I and II involve four clients, while Cases III and IV involve six clients.

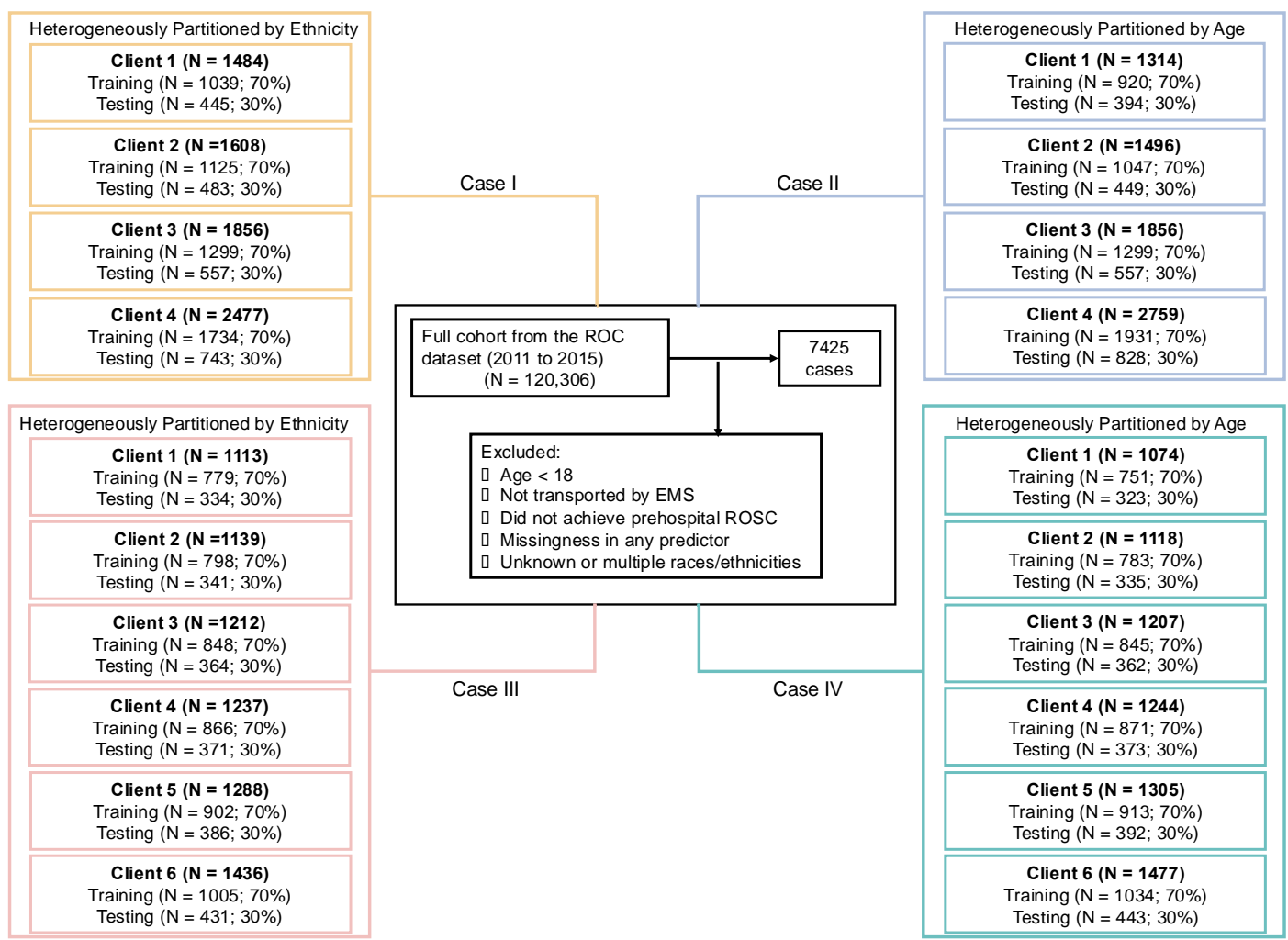

**Figure 3**. Performance comparison of the proposed FairFML method with baseline models, using gender as the sensitive variable. AUC values measure predictive performance, while demographic parity difference (DPD), equalized odds difference (EOD), demographic parity ratio (DPR), and equalized odds ratio (EOR) assess fairness. The horizontal dashed line shows the centralized model's performance, and percentages above the bars indicate relative changes in fairness compared to the centralized model.

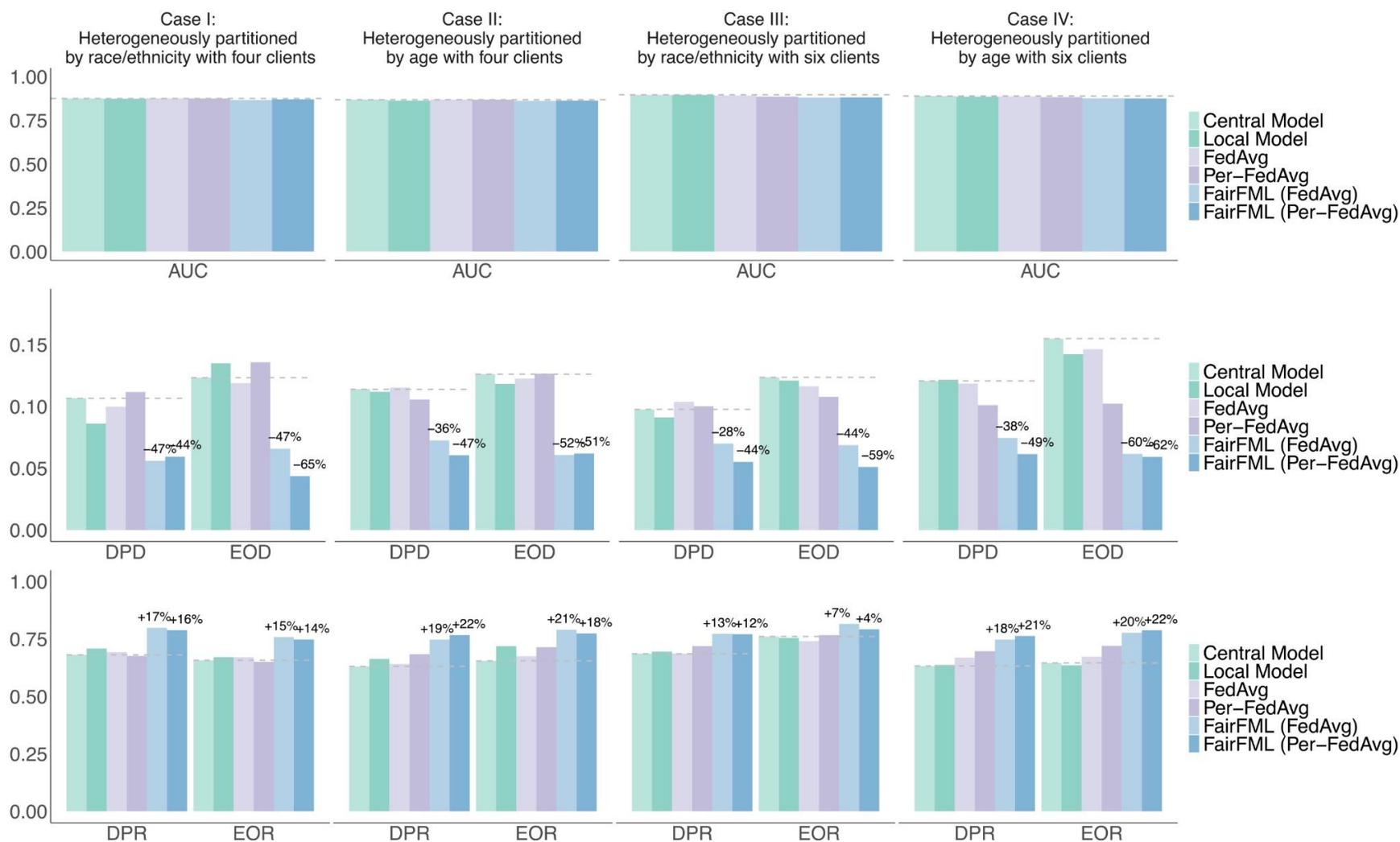

# Supplementary Materials

**eTable 1.** Description of the study cohorts (N=7425).

**eTable 2.** Experimental details of federated learning models training.

**eTable 3.** Detailed results for all four experimental cases.

**eTable 4.** Algorithmic fairness results for gender disparities within each race/ethnicity group.

**eFigure 1.** Pseudocode of FairFML (FedAvg) and FairFML (Per-FedAvg).

**eFigure 2.** λ selection plots using accuracy for all four experimental cases.

**eTable 1.** Description of the study cohorts (N=7425). Data are presented as count (percentage) of patients unless otherwise indicated. Case I: Heterogeneously partitioned by race/ethnicity; four clients. Case II: Heterogeneously partitioned by age; four clients. Case III: Heterogeneously partitioned by race/ethnicity; six clients. Case IV: Heterogeneously partitioned by age; six clients.

| Settings | Client | N | Neurological Outcome (Good) | Age, mean (SD) | Gender (Male) | Etiology of arrest (Cardiac) | Witness Presence (Yes) | Initial rhythm (Shockable) | Bystander CPR (Yes) | Response time, mean (SD) | Adrenaline used (Yes) | Race/Ethnicity | | | |
|---|---|---|---|---|---|---|---|---|---|---|---|---|---|---|---|
| | | | | | | | | | | | | White | Asian | Black | Hispanic |
| Overall | | 7425 | 793 (10.7) | 64.28 (16.30) | 4407 (59.4) | 6835 (92.1) | 4684 (63.1) | 2347 (31.6) | 3495 (47.1) | 5.31 (2.44) | 6401 (86.2) | 4945 (66.6) | 421 (5.7) | 1562 (21.0) | 497 (6.7) |
| Case I | Client 1 | 1484 | 153 (10.3) | 64.63 (16.40) | 896 (60.4) | 1350 (91.0) | 929 (62.6) | 474 (31.9) | 728 (49.1) | 5.37 (2.59) | 1270 (85.6) | 1236 (83.3) | 40 (2.7) | 151 (10.2) | 57 (3.8) |
| | Client 2 | 1608 | 182 (11.3) | 64.53 (16.53) | 945 (58.8) | 1462 (90.9) | 1020 (63.4) | 527 (32.8) | 791 (49.2) | 5.31 (2.39) | 1398 (86.9) | 1236 (76.9) | 48 (3.0) | 256 (15.9) | 68 (4.2) |
| | Client 3 | 1856 | 207 (11.2) | 64.55 (15.82) | 1122 (60.5) | 1715 (92.4) | 1182 (63.7) | 625 (33.7) | 893 (48.1) | 5.31 (2.36) | 1568 (84.5) | 1236 (66.6) | 122 (6.6) | 367 (19.8) | 131 (7.1) |
| | Client 4 | 2477 | 251 (10.1) | 63.71 (16.42) | 1444 (58.3) | 2308 (93.2) | 1553 (62.7) | 721 (29.1) | 1083 (43.7) | 5.28 (2.44) | 2165 (87.4) | 1237 (49.9) | 211 (8.5) | 788 (31.8) | 241 (9.7) |
| Case II | Client 1 | 1314 | 98 (7.5) | 69.89 (14.35) | 746 (56.8) | 1225 (93.2) | 837 (63.7) | 372 (28.3) | 593 (45.1) | 5.38 (2.32) | 1136 (86.5) | 897 (68.3) | 79 (6.0) | 256 (19.5) | 82 (6.2) |
| | Client 2 | 1496 | 144 (9.6) | 67.74 (15.08) | 902 (60.3) | 1384 (92.5) | 965 (64.5) | 447 (29.9) | 679 (45.4) | 5.29 (2.42) | 1304 (87.2) | 1031 (68.9) | 81 (5.4) | 285 (19.1) | 99 (6.6) |
| | Client 3 | 1856 | 206 (11.1) | 64.25 (16.94) | 1119 (60.3) | 1707 (92.0) | 1148 (61.9) | 571 (30.8) | 878 (47.3) | 5.41 (2.57) | 1603 (86.4) | 1228 (66.2) | 119 (6.4) | 376 (20.3) | 133 (7.2) |
| | Client 4 | 2759 | 345 (12.5) | 59.76 (16.08) | 1640 (59.4) | 2519 (91.3) | 1734 (62.8) | 957 (34.7) | 1345 (48.7) | 5.23 (2.42) | 2358 (85.5) | 1789 (64.8) | 142 (5.1) | 645 (23.4) | 183 (6.6) |
| Case III | Client 1 | 1113 | 117 (10.5) | 65.60 (16.13) | 652 (58.6) | 1035 (93.0) | 712 (64.0) | 336 (30.2) | 539 (48.4) | 5.40 (2.50) | 954 (85.7) | 989 (88.9) | 21 (1.9) | 76 (6.8) | 27 (2.4) |
| | Client 2 | 1139 | 134 (11.8) | 64.41 (16.51) | 695 (61.0) | 1050 (92.2) | 719 (63.1) | 378 (33.2) | 552 (48.5) | 5.42 (2.59) | 966 (84.8) | 891 (78.2) | 34 (3.0) | 159 (14.0) | 55 (4.8) |
| | Client 3 | 1212 | 135 (11.1) | 64.84 (15.67) | 714 (58.9) | 1113 (91.8) | 754 (62.2) | 389 (32.1) | 564 (46.5) | 5.28 (2.32) | 1029 (84.9) | 840 (69.3) | 57 (4.7) | 239 (19.7) | 76 (6.3) |
| | Client 4 | 1237 | 136 (11.0) | 64.15 (16.26) | 743 (60.1) | 1122 (90.7) | 773 (62.5) | 402 (32.5) | 600 (48.5) | 5.31 (2.36) | 1084 (87.6) | 791 (63.9) | 88 (7.1) | 270 (21.8) | 88 (7.1) |
| | Client 5 | 1288 | 137 (10.6) | 63.79 (16.49) | 759 (58.9) | 1194 (92.7) | 829 (64.4) | 424 (32.9) | 601 (46.7) | 5.27 (2.29) | 1117 (86.7) | 742 (57.6) | 95 (7.4) | 361 (28.0) | 90 (7.0) |
| | Client 6 | 1436 | 134 (9.3) | 63.25 (16.57) | 844 (58.8) | 1321 (92.0) | 897 (62.5) | 418 (29.1) | 639 (44.5) | 5.24 (2.57) | 1251 (87.1) | 692 (48.2) | 126 (8.8) | 457 (31.8) | 161 (11.2) |
| Case IV | Client 1 | 1074 | 85 (7.9) | 70.71 (12.91) | 652 (60.7) | 1016 (94.6) | 674 (62.8) | 338 (31.5) | 513 (47.8) | 5.33 (2.32) | 949 (88.4) | 747 (69.6) | 54 (5.0) | 207 (19.3) | 66 (6.1) |
| | Client 2 | 1118 | 121 (10.8) | 67.81 (14.97) | 654 (58.5) | 1042 (93.2) | 712 (63.7) | 348 (31.1) | 519 (46.4) | 5.38 (2.64) | 969 (86.7) | 767 (68.6) | 53 (4.7) | 225 (20.1) | 73 (6.5) |
| | Client 3 | 1207 | 104 (8.6) | 65.35 (16.28) | 718 (59.5) | 1102 (91.3) | 747 (61.9) | 359 (29.7) | 553 (45.8) | 5.26 (2.45) | 1058 (87.7) | 791 (65.5) | 79 (6.5) | 259 (21.5) | 78 (6.5) |
| | Client 4 | 1244 | 144 (11.6) | 63.75 (16.61) | 765 (61.5) | 1152 (92.6) | 807 (64.9) | 414 (33.3) | 574 (46.1) | 5.20 (2.18) | 1053 (84.6) | 827 (66.5) | 76 (6.1) | 254 (20.4) | 87 (7.0) |
| | Client 5 | 1305 | 165 (12.6) | 61.51 (16.58) | 755 (57.9) | 1194 (91.5) | 833 (63.8) | 412 (31.6) | 610 (46.7) | 5.37 (2.55) | 1120 (85.8) | 860 (65.9) | 84 (6.4) | 276 (21.1) | 85 (6.5) |
| | Client 6 | 1477 | 174 (11.8) | 58.97 (16.66) | 863 (58.4) | 1329 (90.0) | 911 (61.7) | 476 (32.2) | 726 (49.2) | 5.35 (2.48) | 1252 (84.8) | 953 (64.5) | 75 (5.1) | 341 (23.1) | 108 (7.3) |

**eTable 2.** Experimental details of federated learning models training. For Per-FedAvg, the number of client-side SGD steps is set to 1. All FL models are trained using a local learning rate of 0.1, a batch size of 128, and 10 global iterations. Both $m$ and $m'$ are fixed at 10 across all configurations. The $\gamma$ fine-tuning in step 1 is performed over the range [0.0001,0.1].

| Experiment Settings | Model | $\lambda$ | $\gamma$ fine-tuning range (step 2) | $\gamma$ |
|---|---|---|---|---|
| Case I | FairFML (FedAvg) | 1.0 | [0.0112, 0.0223] | 0.019833 |
| | FairFML (Per-FedAvg) | 1.0 | [0.0112, 0.0223] | 0.0223 |
| Case II | FairFML (FedAvg) | 2.0 | [0.0112, 0.0223] | 0.0112 |
| | FairFML (Per-FedAvg) | 3.0 | [0.0001, 0.0112] | 0.0038 |
| Case III | FairFML (FedAvg) | 2.0 | [0.0001, 0.0112] | 0.008733 |
| | FairFML (Per-FedAvg) | 3.0 | [0.0001, 0.0112] | 0.0075 |
| Case IV | FairFML (FedAvg) | 2.5 | [0.0001, 0.0112] | 0.009967 |
| | FairFML (Per-FedAvg) | 3.0 | [0.0001, 0.0112] | 0.008733 |

**eTable 3.** Detailed results for all four experimental cases.

(a) Case I: Heterogeneously partitioned by race/ethnicity, four clients.

| Testing Data | Model | Prediction Metrics | Fairness Metrics | | | |
|---|---|---|---|---|---|---|
| | | AUC ↑ | DPD ↓ | DPR ↑ | EOD ↓ | EOR ↑ |
| Client 1 | Central Model | 0.8820 | 0.1417 | 0.5811 | 0.0925 | 0.6315 |
| | Local Model | 0.8770 | 0.1034 | 0.6486 | 0.0781 | 0.7165 |
| | FedAvg | 0.8840 | 0.1207 | 0.6393 | 0.0700 | 0.7165 |
| | Per-FedAvg | 0.8823 | 0.1244 | 0.6323 | 0.0742 | 0.7043 |
| | FairFML (FedAvg) | 0.8772 | 0.0799 | 0.7261 | 0.0598 | 0.7652 |
| | FairFML (Per-FedAvg) | 0.8776 | 0.0756 | 0.7336 | 0.0331 | 0.7678 |
| Client 2 | Central Model | 0.8845 | 0.0530 | 0.8403 | 0.0858 | 0.6806 |
| | Local Model | 0.8878 | 0.0502 | 0.8361 | 0.0784 | 0.6730 |
| | FedAvg | 0.8853 | 0.0633 | 0.8176 | 0.1077 | 0.6223 |
| | Per-FedAvg | 0.8857 | 0.0708 | 0.8004 | 0.1160 | 0.6047 |
| | FairFML (FedAvg) | 0.8615 | 0.0286 | 0.8904 | 0.0750 | 0.6831 |
| | FairFML (Per-FedAvg) | 0.8736 | 0.0456 | 0.8612 | 0.0338 | 0.7328 |
| Client 3 | Central Model | 0.8225 | 0.1166 | 0.6639 | 0.2197 | 0.7155 |
| | Local Model | 0.8205 | 0.0901 | 0.6875 | 0.3081 | 0.6433 |
| | FedAvg | 0.8249 | 0.0956 | 0.6845 | 0.1970 | 0.7552 |
| | Per-FedAvg | 0.8270 | 0.1011 | 0.6819 | 0.2197 | 0.7521 |
| | FairFML (FedAvg) | 0.8316 | 0.0676 | 0.8035 | 0.0732 | 0.8526 |
| | FairFML (Per-FedAvg) | 0.8324 | 0.0770 | 0.8026 | 0.0337 | 0.7519 |
| Client 4 | Central Model | 0.9116 | 0.1150 | 0.6401 | 0.0952 | 0.6026 |
| | Local Model | 0.9094 | 0.1008 | 0.6627 | 0.0753 | 0.6516 |
| | FedAvg | 0.9095 | 0.1196 | 0.6310 | 0.1004 | 0.5898 |
| | Per-FedAvg | 0.9054 | 0.1503 | 0.5914 | 0.1332 | 0.5433 |
| | FairFML (FedAvg) | 0.8958 | 0.0481 | 0.7760 | 0.0553 | 0.7336 |
| | FairFML (Per-FedAvg) | 0.8979 | 0.0393 | 0.7560 | 0.0739 | 0.7394 |
| Average | Central Model | 0.8752 | 0.1066 | 0.6814 | 0.1233 | 0.6576 |
| | Local Model | 0.8737 | 0.0861 | 0.7087 | 0.1350 | 0.6711 |
| | FedAvg | 0.8759 | 0.0998 | 0.6931 | 0.1188 | 0.6710 |
| | Per-FedAvg | 0.8751 | 0.1117 | 0.6765 | 0.1358 | 0.6511 |
| | FairFML (FedAvg) | 0.8665 | 0.0561 | 0.7990 | 0.0659 | 0.7586 |
| | FairFML (Per-FedAvg) | 0.8704 | 0.0594 | 0.7884 | 0.0436 | 0.7480 |

(b) Case II: Heterogeneously partitioned by age, four clients.

| Testing Data | Model | Prediction Metrics | Fairness Metrics | | | |
|---|---|---|---|---|---|---|
| | | AUC ↑ | DPD ↓ | DPR ↑ | EOD ↓ | EOR ↑ |
| Client 1 | Central Model | 0.9259 | 0.1485 | 0.4111 | 0.1126 | 0.3835 |
| | Local Model | 0.9148 | 0.1179 | 0.6233 | 0.0844 | 0.6624 |
| | FedAvg | 0.9256 | 0.1407 | 0.4516 | 0.1047 | 0.4416 |
| | Per-FedAvg | 0.9223 | 0.1180 | 0.5826 | 0.0974 | 0.6323 |
| | FairFML (FedAvg) | 0.9198 | 0.0694 | 0.7275 | 0.0308 | 0.7611 |
| | FairFML (Per-FedAvg) | 0.9214 | 0.0309 | 0.7360 | 0.0222 | 0.7631 |
| Client 2 | Central Model | 0.8551 | 0.0691 | 0.7590 | 0.1351 | 0.8649 |
| | Local Model | 0.8496 | 0.0926 | 0.6770 | 0.1351 | 0.8300 |
| | FedAvg | 0.8562 | 0.0711 | 0.7319 | 0.1622 | 0.8378 |
| | Per-FedAvg | 0.8556 | 0.0783 | 0.7126 | 0.1622 | 0.8378 |
| | FairFML (FedAvg) | 0.8449 | 0.0352 | 0.7675 | 0.0733 | 0.8654 |
| | FairFML (Per-FedAvg) | 0.8490 | 0.0722 | 0.7857 | 0.0665 | 0.8649 |
| Client 3 | Central Model | 0.8674 | 0.0456 | 0.8419 | 0.0942 | 0.8934 |
| | Local Model | 0.8637 | 0.0497 | 0.8328 | 0.0942 | 0.8934 |
| | FedAvg | 0.8684 | 0.0404 | 0.8643 | 0.0416 | 0.9375 |
| | Per-FedAvg | 0.8677 | 0.0182 | 0.9401 | 0.0649 | 0.9285 |
| | FairFML (FedAvg) | 0.8599 | 0.0276 | 0.8925 | 0.0056 | 0.9687 |
| | FairFML (Per-FedAvg) | 0.8605 | 0.0164 | 0.9503 | 0.0648 | 0.9285 |
| Client 4 | Central Model | 0.8249 | 0.1924 | 0.5110 | 0.1625 | 0.4783 |
| | Local Model | 0.8249 | 0.1875 | 0.5208 | 0.1594 | 0.4881 |
| | FedAvg | 0.8263 | 0.2091 | 0.5162 | 0.1817 | 0.4837 |
| | Per-FedAvg | 0.8272 | 0.2078 | 0.5018 | 0.1817 | 0.4604 |
|  | FairFML (FedAvg) | 0.8239 | 0.1582 | 0.6037 | 0.1335 | 0.5661 |
|  | FairFML (Per-FedAvg) | 0.8227 | 0.1226 | 0.5981 | 0.0943 | 0.5415 |
| Average | Central Model | 0.8683 | 0.1139 | 0.6308 | 0.1261 | 0.6550 |
|  | Local Model | 0.8633 | 0.1119 | 0.6635 | 0.1183 | 0.7185 |
|  | FedAvg | 0.8691 | 0.1153 | 0.6410 | 0.1226 | 0.6752 |
|  | Per-FedAvg | 0.8682 | 0.1056 | 0.6843 | 0.1266 | 0.7148 |
|  | FairFML (FedAvg) | 0.8621 | 0.0726 | 0.7478 | 0.0608 | 0.7903 |
|  | FairFML (Per-FedAvg) | 0.8634 | 0.0605 | 0.7675 | 0.0620 | 0.7745 |

(c) Case III: Heterogeneously partitioned by race/ethnicity, six clients.

| Testing Data | Model | Prediction Metrics | Fairness Metrics | | | |
|---|---|---|---|---|---|---|
| | | AUC ↑ | DPD ↓ | DPR ↑ | EOD ↓ | EOR ↑ |
| Client 1 | Central Model | 0.9323 | 0.1061 | 0.6451 | 0.1049 | 0.5124 |
| | Local Model | 0.9349 | 0.0895 | 0.6662 | 0.0779 | 0.5679 |
| | FedAvg | 0.9293 | 0.0898 | 0.7046 | 0.0871 | 0.6059 |
| | Per-FedAvg | 0.9191 | 0.1231 | 0.6636 | 0.1253 | 0.5688 |
| | FairFML (FedAvg) | 0.9129 | 0.0663 | 0.7712 | 0.0536 | 0.7177 |
| | FairFML (Per-FedAvg) | 0.9144 | 0.0629 | 0.7254 | 0.0638 | 0.6415 |
| Client 2 | Central Model | 0.9053 | 0.1024 | 0.6788 | 0.0968 | 0.7411 |
| | Local Model | 0.9061 | 0.0906 | 0.6976 | 0.1613 | 0.7431 |
| | FedAvg | 0.8975 | 0.1165 | 0.6651 | 0.0883 | 0.7204 |
| | Per-FedAvg | 0.8890 | 0.0959 | 0.7423 | 0.0645 | 0.7213 |
| | FairFML (FedAvg) | 0.8863 | 0.0810 | 0.6991 | 0.0830 | 0.7652 |
| | FairFML (Per-FedAvg) | 0.8870 | 0.0462 | 0.7530 | 0.0567 | 0.7920 |
| Client 3 | Central Model | 0.8742 | 0.1436 | 0.5873 | 0.2476 | 0.7292 |
| | Local Model | 0.8723 | 0.1318 | 0.5901 | 0.2190 | 0.7527 |
| | FedAvg | 0.8694 | 0.1713 | 0.4949 | 0.2190 | 0.6054 |
| | Per-FedAvg | 0.8791 | 0.1306 | 0.6341 | 0.1535 | 0.7621 |
| | FairFML (FedAvg) | 0.8612 | 0.0982 | 0.6648 | 0.1480 | 0.7690 |
| | FairFML (Per-FedAvg) | 0.8763 | 0.0809 | 0.6881 | 0.0835 | 0.7328 |
| Client 4 | Central Model | 0.8741 | 0.0651 | 0.7789 | 0.0139 | 0.9838 |
| | Local Model | 0.8744 | 0.0655 | 0.7624 | 0.0972 | 0.8889 |
| | FedAvg | 0.8710 | 0.0812 | 0.7150 | 0.0972 | 0.8889 |
|  | Per-FedAvg | 0.8640 | 0.0919 | 0.7022 | 0.0972 | 0.8738 |
|  | FairFML (FedAvg) | 0.8667 | 0.0653 | 0.7762 | 0.0320 | 0.8889 |
|  | FairFML (Per-FedAvg) | 0.8614 | 0.0269 | 0.7713 | 0.0303 | 0.8784 |
| Client 5 | Central Model | 0.8773 | 0.0948 | 0.6647 | 0.2116 | 0.6667 |
|  | Local Model | 0.8784 | 0.0873 | 0.7044 | 0.1032 | 0.6763 |
|  | FedAvg | 0.8775 | 0.0944 | 0.7398 | 0.1058 | 0.7259 |
|  | Per-FedAvg | 0.8689 | 0.0844 | 0.7776 | 0.1058 | 0.7778 |
|  | FairFML (FedAvg) | 0.8624 | 0.0779 | 0.8115 | 0.0406 | 0.8210 |
|  | FairFML (Per-FedAvg) | 0.8614 | 0.0779 | 0.8115 | 0.0389 | 0.8210 |
| Client 6 | Central Model | 0.9124 | 0.0736 | 0.7602 | 0.0667 | 0.9300 |
|  | Local Model | 0.9103 | 0.0814 | 0.7527 | 0.0667 | 0.8998 |
|  | FedAvg | 0.9056 | 0.0699 | 0.7970 | 0.1000 | 0.9000 |
|  | Per-FedAvg | 0.8944 | 0.0754 | 0.7938 | 0.1000 | 0.9000 |
|  | FairFML (FedAvg) | 0.8955 | 0.0307 | 0.9140 | 0.0548 | 0.9333 |
|  | FairFML (Per-FedAvg) | 0.8921 | 0.0360 | 0.8757 | 0.0331 | 0.8889 |
| Average | Central Model | 0.8959 | 0.0976 | 0.6858 | 0.1236 | 0.7605 |
|  | Local Model | 0.8961 | 0.0910 | 0.6956 | 0.1209 | 0.7548 |
|  | FedAvg | 0.8917 | 0.1039 | 0.6861 | 0.1162 | 0.7411 |
|  | Per-FedAvg | 0.8858 | 0.1002 | 0.7189 | 0.1077 | 0.7673 |
|  | FairFML (FedAvg) | 0.8808 | 0.0699 | 0.7728 | 0.0687 | 0.8159 |
|  | FairFML (Per-FedAvg) | 0.8821 | 0.0551 | 0.7708 | 0.0511 | 0.7924 |

(d) Case IV: Heterogeneously partitioned by age, six clients.

| Testing Data | Model | Prediction Metrics | Fairness Metrics | | | |
|---|---|---|---|---|---|---|
| | | AUC ↑ | DPD ↓ | DPR ↑ | EOD ↓ | EOR ↑ |
| Client 1 | Central Model | 0.9180 | 0.1199 | 0.5400 | 0.1304 | 0.6897 |
| | Local Model | 0.9191 | 0.1410 | 0.4238 | 0.1739 | 0.5093 |
| | FedAvg | 0.9167 | 0.1119 | 0.6859 | 0.1394 | 0.7523 |
| | Per-FedAvg | 0.9158 | 0.1245 | 0.6164 | 0.0545 | 0.7692 |
| | FairFML (FedAvg) | 0.9077 | 0.0753 | 0.7157 | 0.0591 | 0.7880 |
| | FairFML (Per-FedAvg) | 0.9057 | 0.0572 | 0.6952 | 0.0476 | 0.7738 |
| Client 2 | Central Model | 0.8969 | 0.0604 | 0.7918 | 0.2615 | 0.7280 |
| | Local Model | 0.8846 | 0.0185 | 0.9383 | 0.1615 | 0.8320 |
| | FedAvg | 0.8968 | 0.0507 | 0.8363 | 0.2000 | 0.8000 |
| | Per-FedAvg | 0.8907 | 0.0237 | 0.9259 | 0.1000 | 0.8806 |
| | FairFML (FedAvg) | 0.8796 | 0.0389 | 0.8889 | 0.0442 | 0.9000 |
| | FairFML (Per-FedAvg) | 0.8780 | 0.0213 | 0.9285 | 0.0465 | 0.9000 |
| Client 3 | Central Model | 0.8619 | 0.2075 | 0.4107 | 0.2424 | 0.4071 |
| | Local Model | 0.8616 | 0.2065 | 0.4217 | 0.2424 | 0.4211 |
| | FedAvg | 0.8601 | 0.2042 | 0.3936 | 0.2424 | 0.3855 |
| | Per-FedAvg | 0.8578 | 0.1750 | 0.4723 | 0.1376 | 0.4679 |
| | FairFML (FedAvg) | 0.8558 | 0.0995 | 0.6255 | 0.0756 | 0.6279 |
| | FairFML (Per-FedAvg) | 0.8544 | 0.0751 | 0.6688 | 0.0779 | 0.7278 |
| Client 4 | Central Model | 0.9157 | 0.1132 | 0.6663 | 0.0858 | 0.6480 |
| | Local Model | 0.9121 | 0.1264 | 0.6414 | 0.0959 | 0.6221 |
| | FedAvg | 0.9100 | 0.1058 | 0.6997 | 0.0784 | 0.6970 |

|  | Method | | | | | |
|---|---|---|---|---|---|---|
|  | Per-FedAvg | 0.9079 | 0.1014 | 0.7085 | 0.1205 | 0.7406 |
|  | FairFML (FedAvg) | 0.8983 | 0.0799 | 0.7865 | 0.0563 | 0.8056 |
|  | FairFML (Per-FedAvg) | 0.9004 | 0.0843 | 0.7774 | 0.0563 | 0.8056 |
| Client 5 | Central Model | 0.8854 | 0.0948 | 0.7299 | 0.1111 | 0.7467 |
|  | Local Model | 0.8779 | 0.1328 | 0.6880 | 0.1015 | 0.6908 |
|  | FedAvg | 0.8781 | 0.1062 | 0.7117 | 0.1111 | 0.7138 |
|  | Per-FedAvg | 0.8709 | 0.0957 | 0.7370 | 0.1389 | 0.7385 |
|  | FairFML (FedAvg) | 0.8642 | 0.0862 | 0.7687 | 0.0831 | 0.7822 |
|  | FairFML (Per-FedAvg) | 0.8680 | 0.0899 | 0.7724 | 0.0854 | 0.7810 |
| Client 6 | Central Model | 0.8592 | 0.1281 | 0.6555 | 0.0979 | 0.6550 |
|  | Local Model | 0.8563 | 0.1049 | 0.7114 | 0.0784 | 0.7378 |
|  | FedAvg | 0.8587 | 0.1321 | 0.6897 | 0.1064 | 0.6884 |
|  | Per-FedAvg | 0.8549 | 0.0860 | 0.7224 | 0.0620 | 0.7255 |
|  | FairFML (FedAvg) | 0.8499 | 0.0669 | 0.7020 | 0.0515 | 0.7597 |
|  | FairFML (Per-FedAvg) | 0.8448 | 0.0407 | 0.7360 | 0.0416 | 0.7402 |
| Average | Central Model | 0.8895 | 0.1207 | 0.6324 | 0.1549 | 0.6458 |
|  | Local Model | 0.8853 | 0.1217 | 0.6374 | 0.1423 | 0.6355 |
|  | FedAvg | 0.8867 | 0.1185 | 0.6695 | 0.1463 | 0.6728 |
|  | Per-FedAvg | 0.8830 | 0.1011 | 0.6971 | 0.1023 | 0.7204 |
|  | FairFML (FedAvg) | 0.8760 | 0.0745 | 0.7479 | 0.0617 | 0.7772 |
|  | FairFML (Per-FedAvg) | 0.8752 | 0.0614 | 0.7631 | 0.0592 | 0.7881 |

**eTable 4.** Algorithmic fairness results for gender disparities within each race/ethnicity group.

| Race | N | Outcome Prevalence | DPD | DPR | EOD | EOR |
|---|---|---|---|---|---|---|
| White | 4945 | 0.1187 | 0.1222 | 0.6458 | 0.0830 | 0.6717 |
| Black | 1562 | 0.0743 | 0.1830 | 0.5027 | 0.1589 | 0.4874 |
| Asian | 421 | 0.1045 | 0.2753 | 0.2626 | 0.3889 | 0.2106 |
| Hispanic | 497 | 0.0926 | 0.0964 | 0.4635 | 0.2500 | 0.5690 |

**eFigure 1.** Pseudocode of FairFML (FedAvg) and FairFML (Per-FedAvg).

---

**Algorithm 1** FairFML (with any selected FL framework $\mathcal{F}$)

---

**Input**: Local epochs $E = [E_1, ..., E_K]$ for $K$ clients, fixed value $\lambda$, initial range of $\gamma$, number of global rounds $T$, batch size $B$, FL framework $\mathcal{F}$
**Output**: Optimal value of $\gamma$: $\gamma_{final}$

Divide the range of $\gamma$ into $m$ equally spaced values: $\Gamma = [\gamma_1, ..., \gamma_m]$
$\gamma_s \leftarrow$ OPTIMIZE_GAMMA($\Gamma, \mathcal{F}$)
Refine the search range around $\gamma_s$ to contain $m'$ equally spaced values: $\Gamma' = [\gamma_{s-1}, ..., \gamma_{s+1}]$
$\gamma_{final} \leftarrow$ OPTIMIZE_GAMMA($\Gamma', \mathcal{F}$)

  **function** OPTIMIZE_GAMMA($\Gamma, \mathcal{F}$)
    **for each candidate value of** $\gamma$ **in** $\Gamma$ **do**
      **for each global round** $t = 1, 2, ..., T$ **do**
        **for each client** $k$ **do**
          Split client $k$'s data into batches of size $B$
          **for each local epoch** $i$ **from** 1 **to** $E_k$ **do**
            Perform client-side SGD update with framework $\mathcal{F}$
            Compute stochastic gradient of loss function with fairness penalty:
            $F_i(w_k^i, D_i) = \mathcal{L}(w_k^i, D_i) + \lambda f(w_k^i, D_i) + \gamma ||w_k^i||_2^2$
          **end for**
          Client $k$ sends updated parameters $w_k$ to the server
        **end for**
        Server updates global model: $w = \frac{1}{K} \sum w_k$
      **end for**
      Evaluate model performance and fairness metrics
    **end for**
    Select $\gamma$ that optimally balances performance and fairness
    **return** optimal $\gamma$
  **end function**

**eFigure 2.** λ selection plots using accuracy for all four experimental cases. First, an unregularized model (i.e., λ =0) is trained as the baseline. Then, λ is incremented in fixed steps (e.g., by 5 in our experiments) until prediction accuracy decreases by more than 0.5% compared to the baseline. For each case, the minimum λ value across all clients is selected.

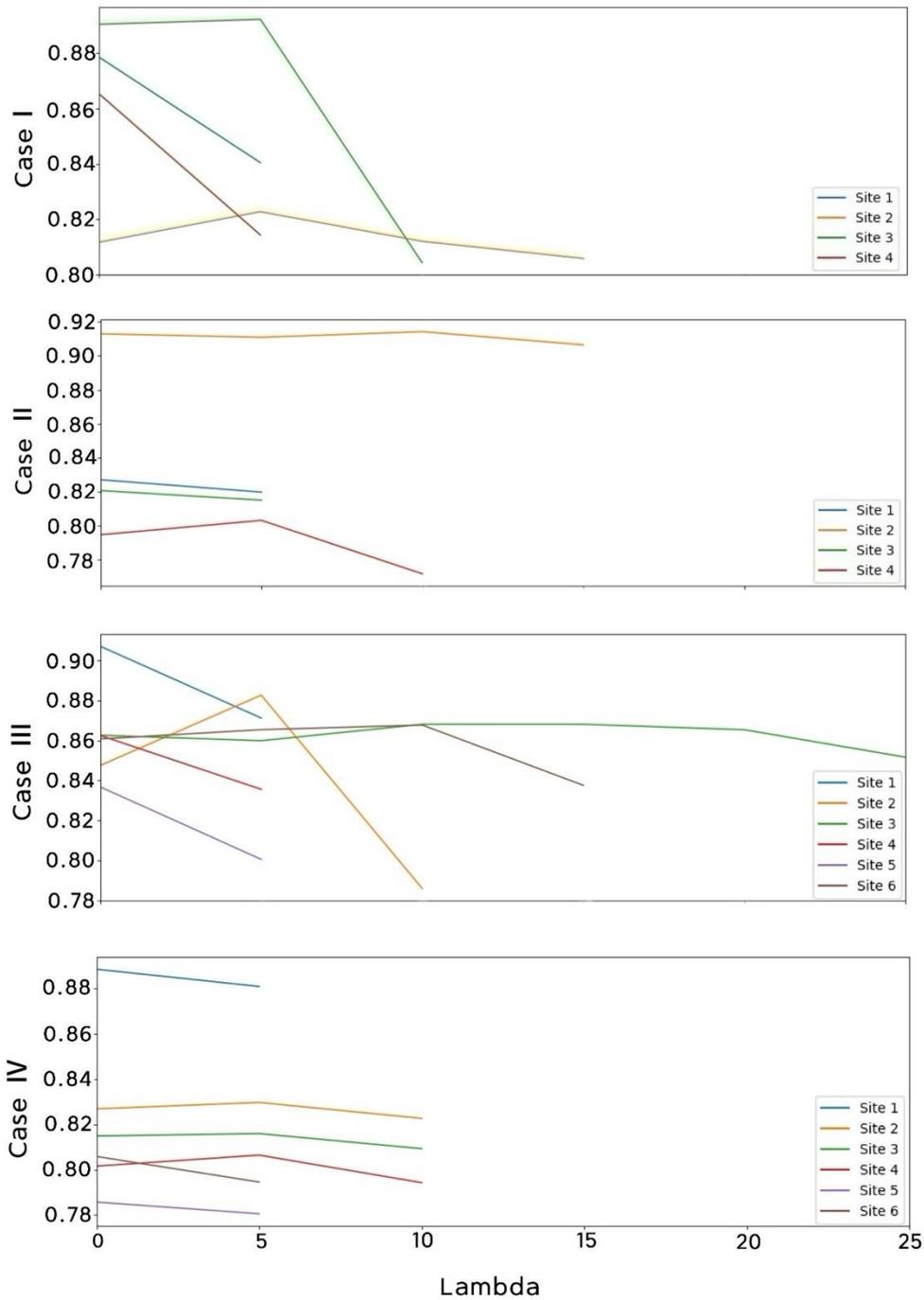